\documentclass[runningheads]{llncs}

\usepackage{multirow}

\usepackage[T1]{fontenc}
\usepackage{graphicx}
\usepackage{hyperref}
\usepackage{color}

\begin{document}
\title{User-to-PC Authentication Through Confirmation on Mobile Devices: On Usability and Performance}
\titlerunning{User-to-PC Authentication Through Confirmation on Mobile Devices}
%
\author{Andreas Pramendorfer\inst{1,2} \and
Rainhard Dieter Findling\inst{1,3} }
\authorrunning{A.\ Pramendorfer \and R.\ D.\ Findling}
%
\institute{SAIL Department, University of Applied Sciences Upper Austria
\and x-tention Informationstechnologie GmbH
\and Google LLC}

\maketitle              
\begin{abstract}

Protecting personal computers (PCs) from unauthorized access typically relies on password authentication, which is know to suffer from cognitive burden and weak credentials. As many users nowadays carry mobile devices with advanced security features throughout their day, there is an opportunity to leverage these devices to improve authentication to PCs. In this paper we utilize a token-based passwordless approach where users authenticate to their PC by confirming the authentication request on their smartphones or smartwatches. Upon a request to login to the PC, or to evaluate privileges, the PC issues an authentication request that users receive on their mobile devices, where users can confirm or deny the request. We evaluate button tap and biometric fingerprint verification as confirmation variants, and compare their authentication duration, success rate, and usability to traditional password-based authentication in a user study with 30 participants and a total of 1,200 authentication attempts. Smartwatch-based authentication outperformed password-based authentication and smartphone-based variants in authentication duration,
while showing comparable success rates.
Participants rated smartwatch-based authentication highest in usability,
followed by password-based authentication
and smartphone-based authentication.

\end{abstract}
\section{Introduction}
Personal computers (PCs) provide access to a noteworthy amount of sensitive data and services that need protection from unauthorized access. Users authenticating to those devices before they can use them, or before they can access sensitive data or services on those devices, is a key part of this protection. However, user authentication mechanisms can bring usability drawbacks~\cite{10.1145/1132736.1132768}. PCs predominantly rely on passwords for user authentication, which are known to suffer from high cognitive load on users and bad scalability across accounts~\cite{6234436}. These drawbacks often cause users to reuse passwords or choose weak passwords~\cite{6234436}. Biometric authentication can offer improved usability over passwords, as users do not have to remember and cannot lose them. However, biometric authentication is unavailable on many desktop computers, and also on a noteworthy portion of consumer laptops.

Nowadays, many users carry one or more personal mobile devices on them throughout their day, such as a smartphone or smartwatch. Those mobile devices present an opportunity to enhance authentication to PCs~-- by incorporating the mobile devices into the authentication process. Modern smartphones are often equipped with advanced security features, including hardware-based secure enclaves for biometrics and cryptographic keys, making them well-suited as secure tokens~\cite{keystore,apple_secure_enclave}. As users already carry these mobile devices on them, they can be utilized as hardware authentication tokens for authentication to the user's PCs, without them having to carry any additional hardware. This could be utilized to improve the usability of authentication to the user's PC, while maintaining authentication security on the level of the security of their mobile devices.

In this paper, we present an authentication approach for PCs that incorporates users' mobile devices to provide a user-friendly, secure, token-based authentication experience. When the user encounters a situation in which they have to authenticate to their PC (e.g.\ to login to the computer, or to elevate their privileges), the computer sends an authentication request to the user's mobile devices. Upon confirming the authentication on the mobile device through a button tap or biometric verification, the mobile device sends the confirmation back to the computer to complete the authentication. The user's mobile devices thereby act as authenticators for their PCs. To enroll a mobile device as an authenticator for a PC, the user registers the public key of a public/private key pair generated on the mobile device with the authentication service for the PC.
On an authentication attempt, the PC notifies the mobile device with a cryptographic challenge. On user confirmation, the mobile device utilizes the private key to sign the cryptographic challenge and send the response back, which the PC verifies with the respective public key. This prevents replay attacks also for situations in which the connection between PC and mobile device would be insecure, and maintains authentication security on the level of the security of the mobile device.

To evaluate the approach, we implement it as a mobile device application for Android and Wear OS that communicates with a personal Linux computer over a REST API. We conduct a user study with 30 participants to answer the following research questions (RQ): RQ1: How does the authentication duration and success rate of this multi-device authentication approach compare to password-based authentication? RQ2: How do users perceive the usability of this multi-device authentication approach over password-based authentication?

\section{Related Work}

We review prior work on utilizing mobile devices as authenticators to unlock other devices or services, without the mobile devices being a second factor in multi-factor authentication.
Most prior work in this area focuses on authentication approaches for the world wide web (web-based authentication). Some prior work covers other areas, such as continuous implicit authentication with mobile devices.

The FIDO2 standard provides an open authentication framework that allows users to authenticate through security keys, so-called passkeys, with the aim to make authentication passwordless.
These passkeys can be stored on devices like smartphones, or dedicated authentication hardware tokens such as FIDO2 authenticator keys~\cite{fido2}. Microsoft Windows natively supports FIDO2 keys, and Linux provides support through extension modules. Both operating systems currently support only specific USB-based authenticator keys for system logons. The FIDO2 Web Authentication (WebAuthn) API enables web applications to utilize passwordless authentication, supporting platform authenticators as well as roaming authenticators~\cite{webauthn}. Platform authenticators include built-in biometrics such as a fingerprint sensor built into a laptop. Roaming authenticators are portable hardware devices like USB security keys. Despite the many security benefits that FIDO2-like authentication protocols offer over password-based authentication, like resistance to phishing attacks or elimination of credential reuse, they still suffer from limitations. Lyastani et al.~\cite{2020_GhorbaniLyastani_CONF} investigated factors that could limit the future adoption of FIDO2. Their findings show that users consider FIDO2 passwordless authentication as more usable and acceptable than password-based authentication. However, the fact that the authenticator is yet another piece of hardware that users have to carry on them to allow them to authenticate in their everyday authentication situations, and the fear of losing those authenticators, which would result in the user losing access and the risks of non-authorized parties gaining access, are the main factors that potentially impede the adoption of FIDO2.

Symbolon~\cite{2022_Laing_CONF} presents a multi-device authentication approach that requires at least \( t \) out of \( n \) authenticator devices to be present for the user to be able to perform an authentication. This addresses the risk of unauthorized access from loss or theft of one authenticator, as it is resilient to loss or theft of up to \( t - 1 \) devices. They use threshold cryptography while staying compliant with the FIDO2 API. They have found Symbolon to be more secure than standard, single-factor FIDO2 while also allowing users to manage authenticators on their side without needing to do so on an account-by-account basis. Symbolon, utilizing the FIDO2 API for user authentication, inherits the standard FIDO2 limitation of primarily supporting USB-based authenticator keys for Windows and Linux system logons, due to operating system constraints.

Chaudhari et al.~\cite{Chaudhari2023} provide a detailed survey of existing authentication systems, which they categorize into password-based, multi-factor, and passwordless methods. They evaluate the security, usability, and scalability of mechanisms such as two-factor authentication (2FA), token-based authentication, biometrics, magic links, and FIDO2 protocols. The paper highlights the vulnerabilities of traditional password systems, particularly their susceptibility to phishing and credential theft, and identifies biometric authentication as the most secure form of passwordless authentication. Based on this analysis, they propose a passwordless authentication framework that uses mobile notifications and biometric input on the mobile device for authentication on a web portal. However, implementation details of the proposed approach are omitted, and the system is not evaluated in a user study, limiting conclusions about its practical effectiveness and usability.

Hintze et al.~\cite{2020_Hintze_CONF} focus on continuous implicit authentication with multiple mobile devices, with the aim to improve the usability and security of mobile device authentication. Their approach continuously verifies the user's identity through different implicit modalities without requiring explicit authentication actions from the user. The authors implemented their approach as an Android authentication framework that uses transparent behavioral and physiological biometrics as authentication modalities, like gait, face, voice, and keystroke dynamics, to continuously evaluate the user’s identity. In their evaluation, their approach was able to reduce the need for users to explicitly authenticate by up to 97\%. However, they also find that for the framework to work in a production environment, support from the mobile device operating system vendors would be needed. The approach of this paper differs from this approach by users explicitly confirming authentication attempts on their mobile devices, instead of the approach continuously authenticating users through different modalities.

ShakeUnlock~\cite{findling2014shakeunlock} proposes to transfer the authentication state from one mobile device to another through conjoint shaking. The approach leverages synchronized accelerometer data to determine if two devices, such as a smartwatch and smartphone, are being shaken together by the same user. Evaluation results indicate that conjoint shaking for 2\,s yields a true match rate of 0.795 and a true non-match rate of 0.867, with a mean unlock time of about 2.5\,s. ShakeUnlock shares similarities with our work, as it leverages mobile devices users carry on them throughout their day as authenticators for other devices, and as it deals with explicit authentication through user gestures. Key differences to our work include shaking as authentication modality and the resulting limitation to users authenticating to mobile devices~-- conjoint shaking cannot be applied with PCs, hence is not applicable for user-to-PC authentication.

\medskip
In summary, prior research has investigated how utilizing multiple mobile devices can help improve usability and security in authentication scenarios. FIDO2-based approaches use authentication tokens and APIs such as WebAuthn to replace password-based with passwordless authentication~\cite{fido2,2022_Laing_CONF}. These approaches predominantly focus on web-based authentication and face usability challenges from users having to carry additional hardware authentication tokens, and from the risk of loss or theft of those tokens~\cite{2020_GhorbaniLyastani_CONF}. Symbolon~\cite{2022_Laing_CONF} addresses the fear of imminent unauthorized access following token loss by using threshold cryptography to enhance security, improve recovery processes, and manage authenticators across devices. However, it inherits the FIDO2 limitation of primarily supporting USB-based authenticator keys for Windows and Linux system logons. Implicit authentication with behavioral and physiological biometrics shows promise for user-friendly authentication, but challenges remain in terms of system integration, device support, and user adoption~\cite{2020_Hintze_CONF}. Additionally, prior research on implicit authentication mainly targets continuous rather than explicit authentication, hence addresses a related but different challenge. ShakeUnlock~\cite{findling2014shakeunlock} proposes a form of explicit authentication by transferring authentication states from one mobile device to another by conjoint shaking, which is not applicable for user-to-PC authentication. Users authenticating to PCs via mobile devices they already carry on them remains largely unaddressed, especially with regards to the resulting authentication success rate, authentication duration, and usability, in comparison to password-based authentication that most PCs use as of today.

\section{Approach}
The approach allows users to authenticate to their PCs through a button tap or biometric verification on their mobile devices that they already carry on them throughout their day.  

To enroll a mobile device such as a smartphone or smartwatch as an authenticator for a PC, the smartphone first generates a public/private key pair. The user registers the public key with the authentication service for the PC, so that the mobile device with the matching private key can later confirm the authentication request for this computer. This ensures that only authorized mobile devices can confirm authentication requests for a PC, and prevents replay attacks even in case the connection between PC and mobile device would be insecure.
Device enrollment, particularly the exchange of public keys between devices and services, is a well-established process in modern authentication systems. Solutions such as the WebAuthn API~\cite{webauthn}
or the U2F registration process~\cite{srinivas2015universal} exemplify device enrollment with public keys. Furthermore, this enrollment needs to be done only once per PC and mobile device~-- and hence does not impact the duration, success rate, or usability of authentication attempts that are conducted after enrollment. For this reason, we declare details of the enrollment process to be out of scope of this paper~-- in the remaining paper, we focus on authentication attempts that are conducted after enrollment.

\begin{figure}[tb]
    \centering
    \includegraphics[width=\linewidth]{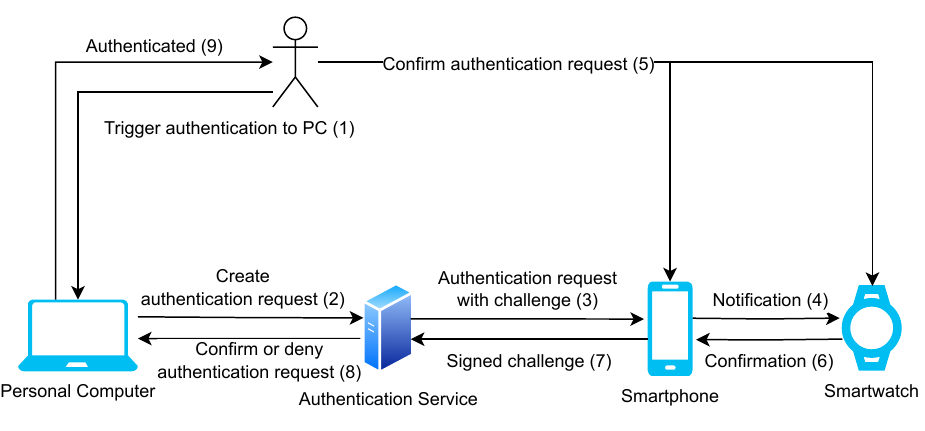} 
    \caption{Overview of the authentication process.}
    \label{fig:flow}
\end{figure}

In authentication scenarios the approach is utilized as follows (figure~\ref{fig:flow}): when the user encounters a situation in which they have to authenticate to their PC (e.g.\ to login to the computer, or to elevate their privileges) they are informed that they can confirm the authentication request on their mobile device, together with a randomly chosen 3-digit comparison code. In the background, the PC sends an authentication request to the authentication service, which creates a cryptographic challenge and sends it to the user's smartphone. On their smartphone, the user is notified of the incoming authentication request via an operating system notification. A tap on this notification opens the authenticator application, where they can confirm or deny the request. There they also see the same randomly chosen 3-digit comparison code, which reassures them they are about to confirm the correct authentication request. We provide and compare different variants for users to confirm their authentication requests on their mobile devices. Those include: to tap an on-screen button on their smartphone or their smartwatch, and to do a biometric fingerprint verification on their smartphone or their smartwatch. Upon user confirmation, the authenticator application signs the cryptographic challenge and sends it back to the authentication service. The authentication service verifies the signature with the public key for the mobile device, and if correct, confirms the authentication attempt.

\subsection{Computer Authentication Module}
The authentication module is integrated with the PC's operating system authentication system. When users attempt to log in to the PC, or to elevate their privileges on the PC, then the operating system authentication system triggers this authentication module as one way to conduct the authentication. The module informs users that they can confirm the authentication on their mobile device, and creates the authentication requests at the authentication service of the approach. The module then waits for the response from the authentication service to accept or deny the authentication attempt. If the response does not arrive within a specific time, the module will deny the authentication attempt.
For our evaluation we implement this module as a PAM (Pluggable Authentication Module). This allows the module to be integrated with the PAM framework, and thereby to authenticate to computers and operating systems that support the PAM framework, such as Linux, without requiring further modifications to the computer, operating system, or any further installed applications.

\subsection{Authentication Service}
The authentication service handles the communication between the authentication module and the mobile device through which users can confirm authentication requests. On an authentication request from the authentication module, it generates a unique challenge and sends it to the mobile device. On receiving the response, it verifies it with the mobile device public key, and notifies the authentication module to accept or deny the authentication request accordingly.
For our evaluation, we implement this authentication service as a REST-API that the mobile applications communicate with,
utilizing the .NET Core framework and its built-in cryptographic features.

\subsection{Mobile Applications}
\label{sec:mobile_applications}
The mobile applications run on users' mobile devices and enable those devices to act as authenticators for the PC. On the first start, the smartphone application generates a public/private key pair and stores the private key in the secure storage of the phone. When the application receives an authentication request, it prompts the user to confirm or deny the authentication. On confirmation it signs the authentication request challenge with the private key and sends the signature back to the authentication service. Different confirmation variants are possible: users can confirm either on their smartphone or smartwatch, either via a button tap or a biometric fingerprint verification. We compare those variants in our evaluation. 
The communication between smartwatch and smartphone is based on the connection between those devices that results from them being paired to each other. 
For our evaluation we implement those mobile applications as Android and Wear OS applications.

\section{Evaluation}
\label{sec:evaluation}
We implement the approach in a prototype\footnote{\url{https://github.com/apramendorfer/mobile-linux-auth}} and evaluate it in a user study with 30 participants and compare its authentication duration, success rate, and usability to password-based authentication.

\subsection{Evaluation Procedure}
The evaluation was conducted in a controlled environment. Participants were seated in an empty room where only the supervisor and the participant were present. A supervisor provided instructions and monitored the session. In the setup phase of the study, participants filled out a pre-study questionnaire that assessed their technical expertise. The questionnaire result allowed the supervisor to assign the participant to either the technically versed or less technically versed group of participants, which had slightly different study procedures (section~\ref{sec:questionnaire}). Then, participants were instructed to choose a password for the PC of the complexity they would typically use for PC authentication, and to not reuse a password they used outside the study. This password was set as the password for the PC. Participants then configured fingerprint authentication on the smartphone, locked it, and stored it as they would when they carry it throughout their day (e.g.\ in their trousers' pocket). They were then instructed to store the smartphone the same way after every authentication they conducted during the study. They then attached the smartwatch, which was already paired with the smartphone, to their preferred wrist. Participants were instructed to unlock the smartwatch and to keep it on their wrist for the remaining study, so that it would not lock from being removed from the wrist. This setup simulates how users typically carry their devices throughout the day: the smartphone locked and stored away, the smartwatch unlocked and worn on their wrist.

After the setup phase, participants engaged in a series of tasks, such as browsing the web or copying text, during which they had to authenticate 8 times to the PC (table~\ref{table:taks}).
Participants repeated this series of tasks 5 times, utilizing a different authentication mechanism for each series: (1) password authentication on the PC, (2) confirmation via button tap on the smartphone, (3) confirmation via biometric verification on the smartphone, (4) confirmation via button tap on the smartwatch, (5) confirmation via biometric verification on the smartwatch. For every participant the first series utilized password authentication, which represents the baseline to compare to. The order of the remaining authentication mechanisms series was chosen randomly for every participant, to avoid potential bias effects from an authentication mechanism coming earlier or later in the study. After completing the 5 series of tasks, participants filled out a usability questionnaire.

\begin{table}[tb]
\centering
\caption{Series of tasks for the less technically versed and technically versed participant groups. Every task that includes login to the PC or execution of a sudo command required authentication.}
\label{table:taks}
\begin{tabular}{|p{0.48\textwidth}|p{0.48\textwidth}|}
\hline
\textbf{Less technically versed} & \textbf{Technically versed} \\
\hline
1. Login to the PC & 1. Login to the PC \\
2. Open a web browser & 2. Open a web browser \\
3. Logout+login to the PC & 3. Logout+login to the PC \\
4. Navigate to a search engine in the web browser & 4. Navigate to a search engine in the web browser \\
5. Logout+login to the PC & 5. Logout+login to the PC  \\
6. Search for an encyclopedia website and open it & 6. Search for an encyclopedia website and open it \\
7. Logout+login to the PC  & 7. Logout+login to the PC  \\
8. Search for "Austria" on the encyclopedia website & 8. Open a terminal emulator \\
9. Logout+login to the PC  & 9. Execute \texttt{sudo apt update} in the terminal emulator \\
10. Scroll down to the "Sports" section & 10. Close and reopen terminal emulator \\
11. Logout+login to the PC  & 11. Execute \texttt{sudo apt install git} in the terminal emulator  \\
12. Open a notepad application & 12. Close and reopen terminal emulator \\
13. Logout+login to the PC  & 13. Execute \texttt{sudo apt remove git} in the terminal emulator  \\
14. Copy a paragraph from the "Sports" section opened earlier into the notepad application & 14. Close and reopen terminal emulator \\
15. Logout+login to the PC  & 15. Execute \texttt{sudo apt autoremove} in the terminal emulator  \\
16. Save the notebook application content into a file on the file system & \\
\hline
\end{tabular}
\end{table}

\subsection{Questionnaires and Measurements}
\label{sec:questionnaire}

In the pre-study questionnaire that participants filled out before the study, they stated their age (numeric), gender (male/female/other), if they currently work in a job that falls into a technical area like IT, engineering, or software development (from "fully disagree" to "fully agree"), if and how often they use a smartphone, smartwatch, Linux, git, or a command line (5 separate questions, "never" to "very often"), as well as their self-assessed level of knowledge in areas such as two-factor authentication, biometric authentication, or token-based authentication (1 question, "no knowledge/no experience" to "expert knowledge/frequent use"). Based on their questionnaire answers, they were assigned to the technically versed or the less technically versed group of participants. Participants that fulfill all of the following were assigned to the technically versed group: they agree or fully agree to work in a technical field, have basic knowledge of authentication methods like two-factor authentication, biometrics, or token-based authentication, and have executed sudo commands on Linux at least rarely.

For authentications that participants performed during the study, we measure the duration that it took participants to conduct the authentication attempt and if the authentication attempt was successful. For unlocking the PC we measure the authentication duration as the time in between participants triggering the PC login screen, which triggers the authentication attempt, and the authentication attempt being confirmed, which unlocks the PC. For sudo commands we measure the authentication duration as the time in between participants submitting the sudo command, which triggers the authentication attempt, and the authentication attempt being confirmed, which executes the sudo command.
We also assess the password complexity of passwords that participants chose for the study via the zxcvbn password strength estimator~\cite{197177}. This yields a score $[0,4]$ per password, from 0 for very weak passwords, to 4 for very strong passwords.

The usability questionnaire that participants filled out after the study contained a total of 22 5-point Likert-scale questions: 2 questions for password-based authentication, to assess participants' perception of ease of use (from "very difficult" to "very easy") and speed (from "very slow" to "very fast"), as the baseline to compare to. And 5 questions for each of the 4 variants of the approach (authentication confirmation via button tap or biometric verification, on smartphone or smartwatch), to assess participants' perception of ease of use and speed (same possible answers as for password-based questions), and also for ease of adaptation (from "very hard" to "very easy"), reliability (from "very unreliable" to "very reliable"), and the likelihood of switching from password-based authentication to the proposed approach if it were available today (from "very unlikely" to "very likely").

\subsection{Participants, Hardware, and Recorded Data}
Participants were recruited for our evaluation study through word of mouth among students of the University of Applied Sciences Upper Austria and their relatives, friends, and colleagues.
Most participants received compensation for their participation in form of a beverage. 30 participants (mean age 34.1 years, age standard deviation 12.3; 17 male, 13 female, 0 other) completed the study. 18 participants were assessed to belong to the technically versed group of participants, 12 to the less technically versed group. While nearly all participants (28) were regular smartphone users, only 10 participants stated that they regularly use a smartwatch. All of the 30 participants completed the series of tasks for each of the 5 authentication variants, which each consist of 8 authentications. This results in 40 authentications per participant, 240 authentications per authentication variant, and 1200 authentications in total. 

All participants used a Google Pixel 6, a Samsung Galaxy Watch4, and a Lenovo Yoga Creator 7 Laptop to complete the evaluation. The Google Pixel 6 comes with an under-display biometric fingerprint sensor which we use for the biometric confirmation. The Galaxy Watch4 does not have a biometric fingerprint sensor. Instead, the mobile application for the evaluation on the smartwatch mocks a biometric fingerprint sensor as follows: it imitates the UI for under-display fingerprint sensors found on Android smartphones by displaying a round button with a fingerprint icon. When pressed, the button triggers a ripple effect, and after being pressed for 400\,ms, the application confirms the authentication. If pressed less than 400\,ms the authentication fails. Participants were informed that the fingerprint sensor on the smartwatch is simulated.

\section{Results and Discussions}
Table~\ref{tab:auth_mechanisms} shows the duration required to perform the authentication and the authentication success rate for all variants of the approach. Authentication variants with confirmation on the smartwatch were the fastest, with a mean authentication duration of 4.5\,s (button tap) and 4.6\,s (biometric verification). Password-based authentication to the PC ranked second, with a mean duration of 4.6\,s. In contrast, smartphone-based authentication confirmation was noticeably slower with a mean duration of 7.0\,s. One factor that causes confirmations to take noticeably longer on the smartphone than on the smartwatch is that users first have to authenticate to the smartphone to unlock it, before they can confirm the authentication request. In contrast, the smartwatch remains unlocked on their wrist, which allows to confirm the authentication request without having to at first unlock the smartwatch. On both the smartphone and smartwatch, confirming via biometric verification was slower, which aligns with biometric verification taking longer than tapping a confirmation button on a device. 

Almost all participants (27) chose a strong or very strong password. As a result, the authentication time measurements for password-based authentication are primarily based on strong and very strong passwords. Due to the small number of weaker passwords in the study, we refrain from drawing conclusions about the relationship between weaker passwords and study results.
Furthermore, all authentication approaches had a high success rate. Password-based authentication achieved a 97\% success rate, and the mean success rate for the approach was 98\%, ranging from 96-100\%. The differences between the confirmation variants were minor (table~\ref{tab:auth_mechanisms}).

\begin{table}[tb]
    \centering
    \caption{Mean authentication duration (standard deviation) and mean authentication success rate, per variant and group of participants.}
    \begin{tabular}{|l|cc|cc|cc|}
        \hline
        \multirow{2}{*}{\textbf{Variant}} & \multicolumn{2}{c|}{\textbf{All}} & \multicolumn{2}{c|}{\textbf{Technical}} & \multicolumn{2}{c|}{\textbf{Less Technical}} \\
        & Duration & Success & Duration & Success & Duration & Success \\
        \hline
        Password & 4.6\,s (1.8\,s) & 97\% & 3.5\,s (0.9\,s) & 97\% & 6.4\,s (1.4\,s) & 97\% \\
        Phone (Button) & 6.6\,s (2.3\,s) & 100\% & 6.5\,s (1.7\,s) & 100\% & 6.7\,s (3.0\,s) & 100\% \\
        Phone (Biometric) & 7.4\,s (2.0\,s) & 98\% & 6.9\,s (1.8\,s) & 100\% & 8.1\,s (2.3\,s) & 96\% \\
        Watch (Button) & 4.5\,s (2.0\,s) & 96\% & 4.4\,s (1.8\,s) & 96\% & 4.6\,s (2.4\,s) & 96\% \\
        Watch (Biometric) & 4.6\,s (1.5\,s) & 98\% & 5.0\,s (1.8\,s) & 97\% & 3.9\,s (1.2\,s) & 99\% \\
        \hline
    \end{tabular}
    \label{tab:auth_mechanisms}
\end{table}

Participants generally found the approach easy to use (table~\ref{tab:usability_results}), with smartwatch-based authentication receiving the highest rating (mean 4.7), followed by password-based authentication (mean 4.5), and smartphone-based authentication (mean 4.0). The inclusion of biometric verification did not noticeably impact how participants perceived ease of use.

\begin{table}[tb]
    \centering
    \caption{Mean results of the usability questionnaire, for all participants (All), technically versed participants (T), and less technically versed participants (LT). Answers reflect users' perception of ease of use (Q1), speed (Q2), ease of adoption (Q3), reliability (Q4), and likelihood of utilizing the approach if it were available today (Q5).}
    \begin{tabular}{|l|ccc|ccc|ccc|ccc|ccc|}
        \hline
        \multirow{2}{*}{\textbf{Variant}} 
        & \multicolumn{3}{c|}{\textbf{Q1}} 
        & \multicolumn{3}{c|}{\textbf{Q2}} 
        & \multicolumn{3}{c|}{\textbf{Q3}} 
        & \multicolumn{3}{c|}{\textbf{Q4}} 
        & \multicolumn{3}{c|}{\textbf{Q5}} \\
        & All & T & LT & All & T & LT & All & T & LT & All & T & LT & All & T & LT \\
        \hline
        Password           & 4.5 & 4.7 & 4.3 & 3.4 & 3.5 & 3.3 & n/a & n/a  & n/a  & n/a  & n/a  & n/a  & n/a  & n/a  & n/a  \\
        Phone (Button)     & 3.9 & 3.7 & 4.3 & 3.7 & 3.5 & 4.0 & 4.1 & 4.0 & 4.3 & 4.1 & 3.8 & 4.5 & 3.2 & 3.3 & 3.0 \\
        Phone (Biometric)  & 4.1 & 4.0 & 4.3 & 3.9 & 3.8 & 4.0 & 3.9 & 3.8 & 4.0 & 4.0 & 3.8 & 4.3 & 3.2 & 3.7 & 2.5 \\
        Watch (Button)     & 4.7 & 4.5 & 4.9 & 4.6 & 4.5 & 4.8 & 4.6 & 4.5 & 4.8 & 4.2 & 4.0 & 4.5 & 3.8 & 4.2 & 3.3 \\
        Watch (Biometric)  & 4.6 & 4.5 & 4.8 & 4.5 & 4.5 & 4.4 & 4.7 & 4.7 & 4.8 & 4.4 & 4.3 & 4.5 & 4.1 & 4.5 & 3.5 \\
        \hline
    \end{tabular}
    \label{tab:usability_results}
\end{table}

When asked if they would use the approach instead of password-based authentication if it were available today, participants expressed stronger interest in the smartwatch-based variants (mean 4.0) compared to smartphone-based variants (mean 3.2). Inclusion of biometric verification had no noticeable impact on users’ interest to use the approach over password-based authentication.
This indicates that the choice of device, which in turn impacts the duration it takes to perform an authentication, likely is a stronger decision factor than the choice between button tap or biometric verification as authentication confirmation method.

In between the technically versed and less technically versed groups of participants we observed a noticeable difference in authentication duration for password-based authentication (table~\ref{tab:auth_mechanisms}). Password-based authentication took noticeably less time for participants in the technically versed group (mean 3.5\,s) over the less technically versed group (mean 6.4\,s). In contrast, for the smartphone- and smartwatch-based approach, this gap in authentication duration was noticeably smaller between the two groups (mean difference of 0.7\,s for both approaches). This indicates that the authentication duration of the approach is less affected by the users' technical expertise than password-based authentication. No noticeably relevant difference in success rates was found between the two groups.

Usability results also varied between the technically versed and the less technically versed group of participants. When asked how likely they would use the approach over password-based authentication if it were available today, technically versed users rated the smartwatch variant with biometric confirmation with a mean of 4.5, while less technically versed users rated it with a mean of 3.5. This difference could indicate that greater familiarity with the technology and the respective mobile devices increases the likelihood of users adopting the approach.

Limitations of the approach and its evaluation include that the security of the approach inherently depends on the security of the mobile devices used as authenticators. If a mobile device used as authenticator gets compromised, then this also compromises the approach and allows attackers to confirm an authentication attempt to the PC. However, this also requires attackers to obtain access to the mobile device that serves as authenticator device~-- either in an already unlocked state, or with the additional effort of also obtaining the means to unlock the mobile device.
Concretely, if a smartphone used as authenticator is protected by a weak authentication mechanism (e.g.\ a weak PIN), if attackers obtain physical access to it, then they might be able to unlock and compromise it. In contrast, smartwatches typically stay unlocked while attached to the user's wrist and lock when being taken off the wrist. For this reason attackers could create a malicious authentication request in a moment where the user does not observe the notification on their smartwatch, e.g.\ because they are distracted. Then the attackers could try to confirm the authentication attempt on the unlocked smartwatch attached to the user's wrist, without the user noticing, e.g.\ because they are in a crowded space.
We leave it to future work to investigate how likely users are to notice such attacks.

Furthermore, password-based authentication to the PC will still need to be available as a fallback authentication for situations in which confirming authentication attempts via a mobile device is not possible. As a result, the security of the PC is still inherently dependent on the strength of its authentication password. However, as users would not need to enter or remember their fallback authentication password on a regular basis, they could choose a significantly stronger password that they look up in the rare cases when they need it. We leave it to future work to investigate the impact that regularly confirming authentication attempts on mobile devices instead of entering a password has on the strength of the password that users would choose for fallback authentication.

\section{Conclusion}
In this paper we investigated how authentication duration, success rate, and usability of users confirming authentication requests to their PCs on mobile devices they possess and already carry throughout their day compare to users authenticating to their PCs via entering a password. For the former, we presented a multi-device authentication approach that allows users to trigger an authentication on their PCs, e.g.\ to unlock it or to elevate their privileges, and then to confirm or deny the authentication request on a mobile device, e.g.\ a smartphone or smartwatch. We investigated different variants for how users can confirm this on their mobile devices: via a button tap or a biometric fingerprint verification, on their smartphone or their smartwatch.

To answer the research questions (RQ) posed in this paper we conducted a user study with 30 participants that evaluates and compares the different confirmation variants with traditional password-based authentication. With regards to RQ1, our results indicate that confirmation on the smartwatch is the fastest variant, followed by password-based authentication and confirmation on the smartphone (4.6\,s, 4.6\,s, and 7.0\,s mean duration to authenticate, respectively). Success rates for confirmation on mobile devices are comparable to password-based authentication (98\% and 97\% mean authentication success rates). With regards to RQ2, participants rated the usability of confirmation on the smartwatch the highest, followed by password-based authentication and confirmation on the smartphone (4.7, 4.5, and 4.0 on a 5-point Likert-scale, respectively). Participants also indicated a stronger interest to switch their PC authentication to confirming on a smartwatch than on a smartphone, if the approach were available today (4.0 and 3.2 on a 5-point Likert-scale).

While our findings highlight the potential to improve user authentication to PCs through confirmation on mobile devices users already carry throughout their day, we acknowledge several limitations of our evaluation. The setup process, including device pairing and key exchange, was not part of the evaluation. These steps could impact user adoption and satisfaction in real-world scenarios. Also, the evaluation does not investigate how easy it would be for attackers to confirm an authentication request on the user's smartwatch while its attached to the user's wrist, without the user noticing. Furthermore, the impact of confirmation fatigue, where users are overwhelmed by the number of confirmations they have to do, and susceptibility to confirmation fatigue attacks, where attackers exploit that users have developed a habit of approving authentication requests on their mobile devices without verifying their legitimacy, have not been assessed.

Future research could investigate how strongly this setup process impacts usability, and how easy it would be for attackers to conduct those attacks. Furthermore, future research could investigate alternative methods of users confirming an authentication request on their mobile device, such as user gestures sensed by smartwatch sensors. It could also investigate user-friendly recovery mechanisms as an alternative to password-based authentication fallback for when the mobile device as the primary authenticator fails or is unavailable, e.g.\ through approaches like secure token syncing or threshold cryptography.

%
%
%
\bibliographystyle{splncs04}
\bibliography{paper}

\end{document}